\documentclass{appolb}
\usepackage{graphicx}  
\usepackage[backend=biber,sorting=none]{biblatex}
\begin{document}
\title{Deep-inelastic scattering with collider neutrinos at the LHC and beyond%
\thanks{Presented at ``Diffraction and Low-$x$ 2024'', Trabia (Palermo, Italy), September 8-14, 2024.}
}
\author{Toni M{\"a}kel{\"a}
\address{Department of Physics and Astronomy, University of California, Irvine, CA 92697 USA}
}
\maketitle
\begin{abstract}
Proton-proton collisions at the LHC generate high-intensity collimated beams of forward neutrinos up to TeV energies. Their recent observations and the initiation of a novel LHC neutrino program motivate investigations of this previously unexploited beam. The kinematic region for neutrino deep-inelastic scattering measurements at the LHC overlaps with that of the Electron-Ion Collider. The effect of the LHC $\nu$DIS data on parton distribution functions (PDFs) is assessed by generating projections for the Run 3 LHC experiments, and for select proposed detectors at the HL-LHC. Estimating their impact in global (n)PDF analyses reveals a significant reduction of PDF uncertainties, particularly for strange and valence quarks. Furthermore, the effect of neutrino flux uncertainties is examined by parametrizing the correlations between a broad selection of neutrino production predictions in forward hadron decays. This allows determination of the highest achievable precision for neutrino observations, and constraining physics within and beyond the Standard Model. This is demonstrated by setting bounds on effective theory operators, and discussing the prospects for an experimental confirmation of the enhanced strangeness scenario proposed to resolve the cosmic ray muon puzzle, using LHC data. Moreover, there is promise for a first measurement of neutrino tridents with a statistical significance exceeding 5$\sigma$.
\end{abstract}
  
\section{Introduction}

The first observations of neutrinos produced at the Large Hadron Collider (LHC) by the FASER~\cite{FASER:2023zcr} and SND@LHC~\cite{SNDLHC:2023pun} Collaborations have initiated the era of accelerator-based neutrino research at the TeV scale. Moreover, FASER has already performed the first measurement of $\nu_e$ and $\nu_\mu$ interaction cross sections~\cite{FASER:2024hoe}. While the current experiments demonstrate great potential, and both FASER~\cite{Boyd:2882503} and SND@LHC~\cite{Abbaneo:2895224, Abbaneo:2909524} intend to continue and upgrade their operations during the LHC Run~4, it is essential for maximizing the physics potential of the high-luminosity LHC run to continue expanding this program. To this end, a purpose-built Forward Physics Facility (FPF) housing several larger experiments has been proposed~\cite{Anchordoqui:2021ghd, Feng:2022inv, Adhikary:2024nlv} at a distance of 620~m along the tangential line of sight from the ATLAS interaction point. The prospects for similar experiments at the Future Circular Collider~\cite{MammenAbraham:2024gun} have also been considered.

The neutrinos observed at such experiments result from the weak decays of hadrons produced in the initial collisions at the LHC. While these, together with potential forward long-lived particles, are never observed in central experiments, the neutrinos may initiate interactions such as deep inelastic scattering (DIS) in the fixed target of a forward detector. The rates of both neutrino production in hadonic decays, and their interactions with the detector, can however be affected by various physics effects. The present work summarizes three publications, Ref.~\cite{Cruz-Martinez:2023sdv, PhysRevD.108.095020, Altmannshofer:2024hqd}, illustrating how LHC neutrinos probe physics both within and beyond the Standard Model (SM). 

The paper is organized as follows. Sec.~\ref{sec:PDFs} discusses probing proton and nuclear structure, while Sec.~\ref{sec:fluxfit} addresses uncertainties in neutrino flux predictions and ultimate constraints achievable at the FPF, and Sec.~\ref{sec:tridents} discusses the prospect of a conclusive observation of neutrino tridents. Conclusions are presented in Sec.~\ref{sec:conclusions}.

\section{Probing proton and nuclear structure with LHC neutrinos}
\label{sec:PDFs}

\begin{figure}[htb]
\centerline{%
\includegraphics[width=0.45\textwidth]{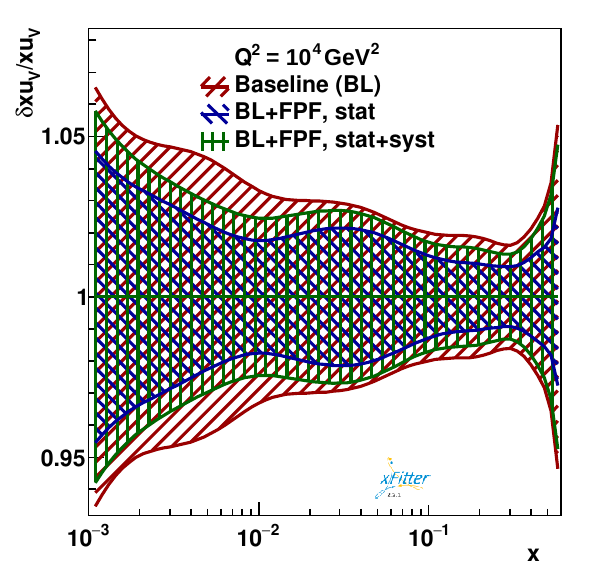}
\includegraphics[width=0.45\textwidth]{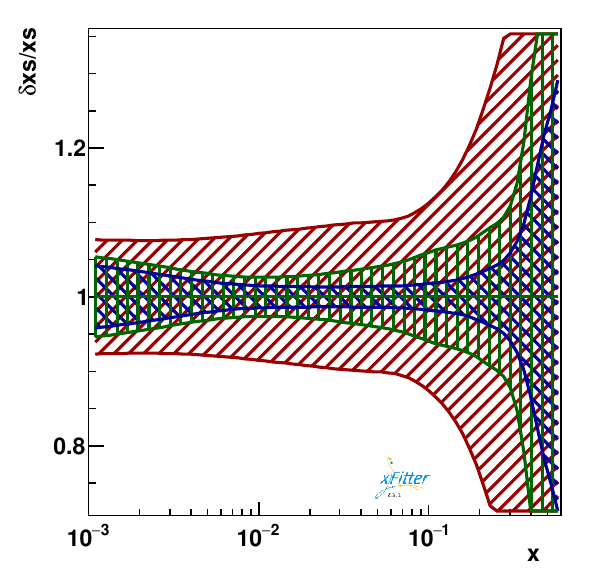}}
\caption{Fractional uncertainties (68\% CL) at $Q^2 = 10^4$~GeV$^2$ for the up valence (left) and strange (right) quarks in the PDF4LHC21 baseline (red), compared to the results of Hessian profiling performed with FPF pseudodata. Projections accounting for estimated statistical(+systematic) uncertainties are shown in blue (green). Taken from Ref.~\cite{Cruz-Martinez:2023sdv}.}
\label{Fig:FPFPDF}
\end{figure}

The impact of the FPF data on global parton distribution functions (PDF) is estimated via a Hessian PDF profiling procedure~\cite{Paukkunen:2014zia, Schmidt:2018hvu, AbdulKhalek:2018rok, HERAFitterdevelopersTeam:2015cre}, implemented in the \textsc{xFitter} open-source QCD analysis framework~\cite{Alekhin:2014irh, Bertone:2017tig, xFitter:2022zjb, xFitter:web}.
Fig.~\ref{Fig:FPFPDF} shows the results using the PDF4LHC21~\cite{PDF4LHCWorkingGroup:2022cjn} proton PDF, assuming an isoscalar free nuclear target. Most improvement is observed for the valence and strange quarks. The improvement in the valence (strange) quark PDFs relies in paricular on the possibility of lepton charge identification (charm tagging). Similar improvement is also observed after accounting for nuclear corrections using the EPPS21~\cite{Eskola:2021nhw} set~\cite{Cruz-Martinez:2023sdv}.

The coverage of charged current (CC) interactions in the $x,Q^2$ kinematic plane expected at the FPF overlaps with that of the Electron-Ion Collider (EIC)~\cite{AbdulKhalek:2021gbh}, providing complementary information to the neutral current (NC) scattering measurements at the EIC~\cite{Cruz-Martinez:2023sdv}. 
Notably, although PDF4LHC21 already includes neutrino DIS data, the FPF is hereby shown to provide even further constraints. LHC run 3 statistics are however determined insufficient for constraining PDFs, thus further motivationg the FPF and EIC. The projected improvement in PDF uncertainties will increase the precision of many SM cross section measurements, relevant to key processes such as inclusive Drell-Yan and measurements of the W boson mass and 
the Weinberg angle~\cite{Cruz-Martinez:2023sdv}.

Moreover, similar work in the context of the FCC has indicated great potential for studying e.g. polarized PDFs and cold nuclear matter in $pPb$ collisions. 
The latter relies on describing charm production perturbatively, and of tying the forward neutrino data to events at the central experiment, probing nuclear PDFs at $x\sim10^{-9}$~\cite{MammenAbraham:2024gun}.

\section{The neutrino spectra and flux uncertainties}
\label{sec:fluxfit}

As forward neutrino experiments are probing a previously unexplored kinematic region, there are large discrepancies between various neutrino flux predictions due to their reliance on different phenomenological model assumptions which affect the shape and magnitude of the $\nu$ spectra. The size of the prediction envelope is very large, and it is crucial to ensure that sought-after physics effects are not obscured by uncertainties. Based on a Fisher information approach, Ref.~\cite{PhysRevD.108.095020} presents a framework for obtaining the smallest uncertainty achievable in a measurement. This is done by parametrizing the correlations in the energy and radial distributions, as well as the neutrino flavor and parent hadron composition between a broad selection of predictions.

\begin{figure}[htb]
\centerline{%
\includegraphics[width=0.67\textwidth,trim={0cm 0cm 12cm 0cm},clip]{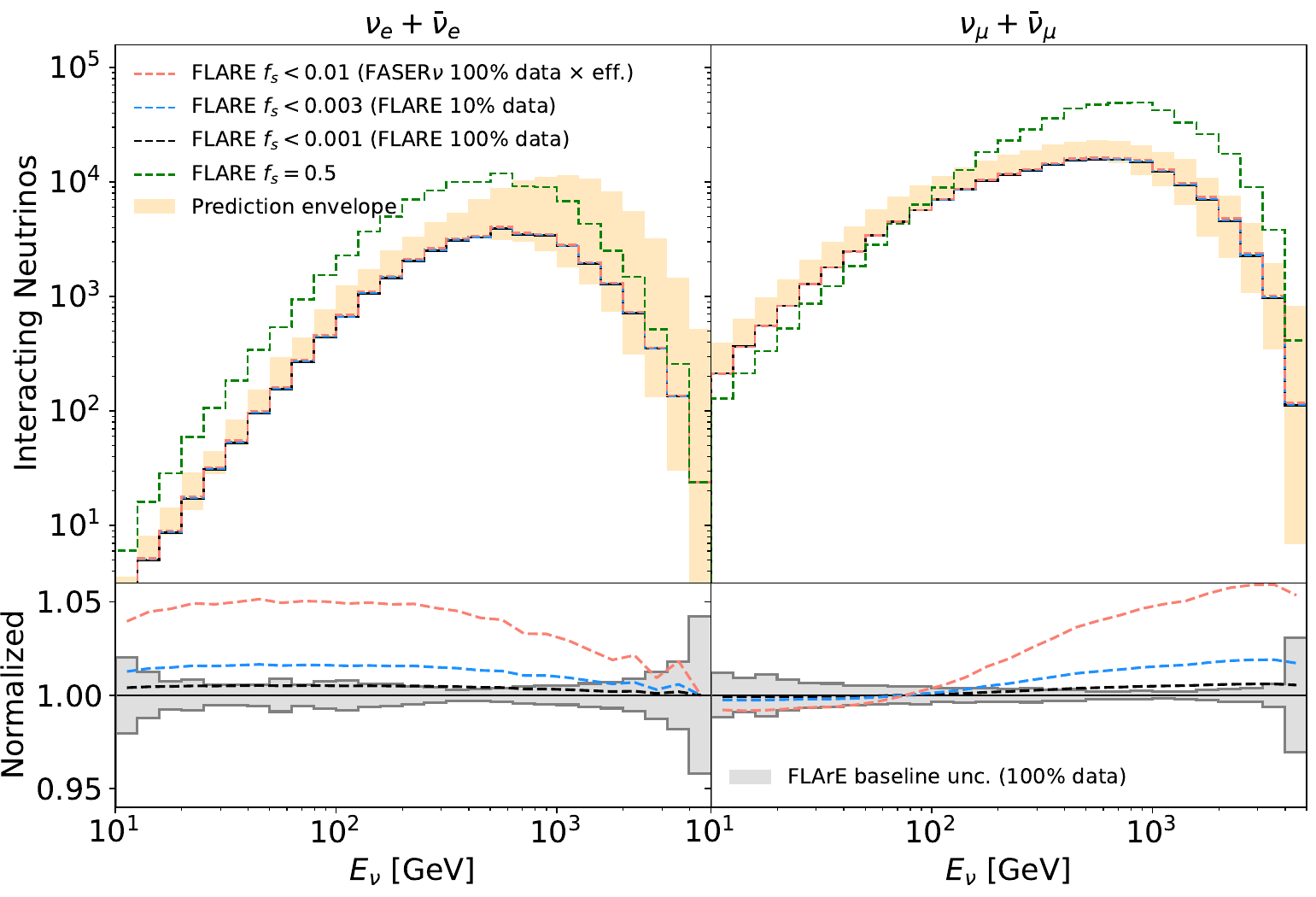}
\includegraphics[width=0.33\textwidth]{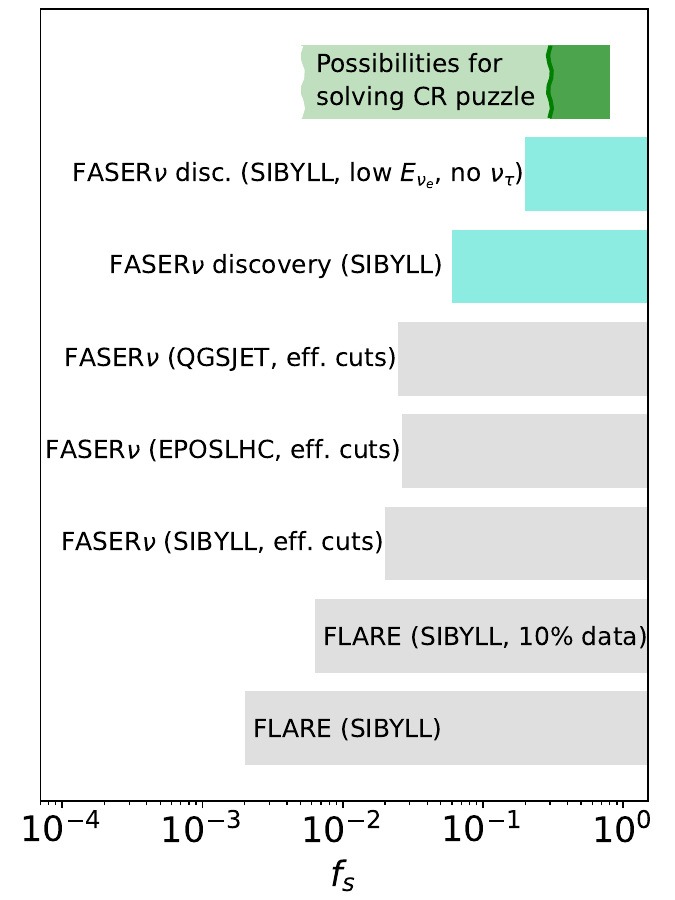}}
\caption{
\textit{Left:} The $\nu_e$ and $\nu_\mu$ CC rates in the FLArE experiment to be hosted at FPF. The solid black line has no strangeness enhancement, while the dashed orange (blue, black) correspond to the $1\sigma$ exclusion bounds obtained for FASER$\nu$ and FLARE with 10\% or 100\% of the expected data, respectively. The $f_s=0.5$ case, which could solve the cosmic-ray muon puzzle, is shown in green. The yellow-shaded band indicates the broad envelope of predictions based on which the much smaller ultimate uncertainty band (gray) is obtained
\textit{Right:} The $2\sigma$ constrained values (gray) for $f_s$ obtained using FLARE and FASER$\nu$, compared to the discovery potential at FASER$\nu$ (turquoise), with and without the information on $\nu_\tau$ and high-energy contributions to the $\nu_e$ spectrum. Notably, all the bounds cover the $0.3 < f_s < 0.8$ region shown in dark green, i.e., the values favored by the enhanced strangeness solution to the CR muon puzzle. The light green band indicates that the effect might however manifest in a more subtle way in $pp$ collisions at the LHC. Taken from Ref.~\cite{PhysRevD.108.095020}.
}
\label{Fig:enhancedStrangeness}
\end{figure}

This allows investigating the most stringent exclusion bounds obtainable for various processes at existing and proposed detectors. The LHC is demonstrated to help for instance in solving the cosmic ray muon puzzle, a significant deficit of high-energy muons in air shower simulations compared to measurements, first observed in the Pierre Auger Observatory data~\cite{PierreAuger:2014ucz,PierreAuger:2016nfk,PierreAuger:2021qsd}. The issue could be caused by a mismodeling of the distribution of produced secondary particles, and the enhanced strangeness hypothesis suggests that the number of muons is increased by enhanced $s$ production, leading to less pions and more kaons at LHC. A phenomenological model, in which the counts of neutrinos originating from pion decays are reweighed by a factor of $(1-f_s)$ and kaons by $(1 + F f_s)$, with $F$ a factor accounting for the difference in $\pi / K$ production rates, could explain the muon puzzle if $f_s=0.5$~\cite{Anchordoqui:2022fpn}. As illustrated in Fig.~\ref{Fig:enhancedStrangeness}, such values can be well constrained by FASER$\nu$ already during run 3. It should be noted that$f_s$ might also have lower values at LHC energies, and such cases could be probed by the FPF.

It is also possible to constrain non-standard interactions (NSI) by extending the SM Lagrangian by dimension-6 effective operators, connecting a $u$ and $d$ quark to a $\nu_\tau$ and a muon or to a $\nu_e$ and a $\tau$~\cite{Falkowski:2021bkq}. The former affects the rate of incoming neutrinos due to modifying pion decays, while the latter affects the rates of interactions observed at the detector. The projected FPF limits are observed to improve on contemporary constraints already with just 10\% of the expected data, while the full result could improve the bounds on specific operators by an order of magnitude~\cite{PhysRevD.108.095020}.

\section{Neutrino tridents at FASER$\nu$2}
\label{sec:tridents}

The experimental observation of neutrino tridents, i.e. the production of a three-lepton final state in neutrino scattering off a nucleus through a photon exchange, has previously been found to be a notoriously difficult task. Detections were claimed at CHARM-II~\cite{CHARM-II:1990dvf} and CCFR~\cite{CCFR:1991lpl}, while NuTeV~\cite{NuTeV:1999wlw} identified diffractive charm production as a previously neglected background and did not claim observation. Therefore, the main task of contemporary work on tridents is to assess the possibility of resolving the signal from the backgrounds, the two main ones at FASER$\nu$2 being single pion and charmed hadron production. 

\begin{figure}[htb]
\centerline{%
\raisebox{3mm}{\includegraphics[width=0.35\textwidth]{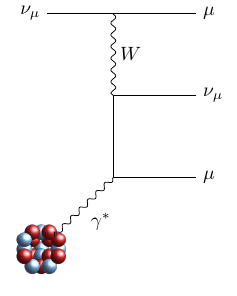}}
\includegraphics[width=0.38\textwidth]{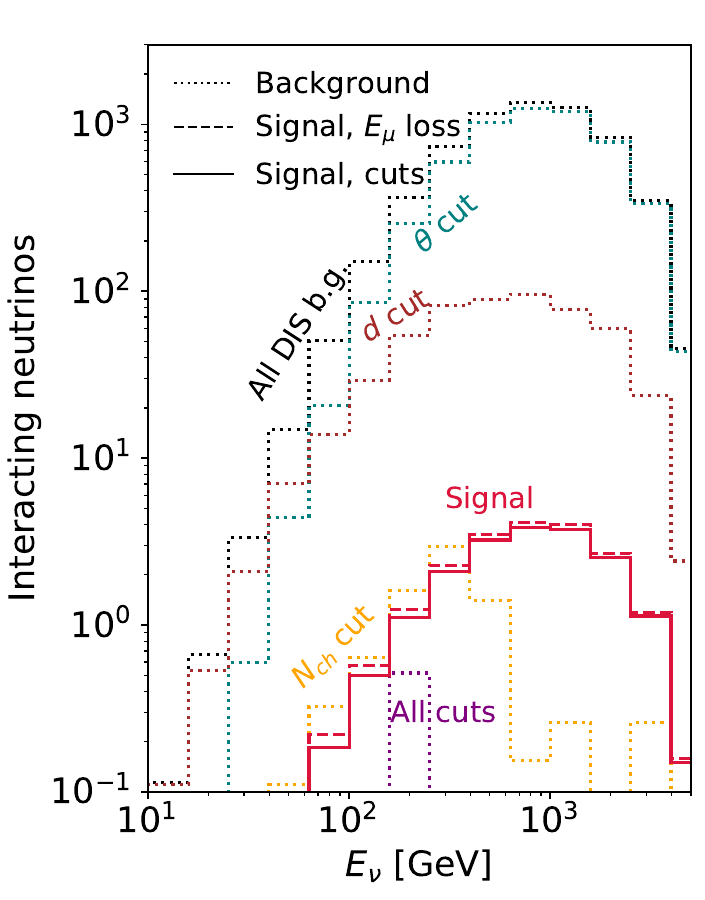}}
\caption{\textit{Left}: A sample Feynman diagram for trident production of a dimuon pair in an interaction with incident muon neutrinos. 
\textit{Right}: The trident signal including only events permitting reverse tracking of both muons is shown in red, with the dashed (solid) line corresponding to before (after) all other cuts employed to reject the background. These account for the reverse tracking requirement that both muons survive an energy loss while traversing the detector matter. The background events surviving each cut are shown as dashed lines. Taken from Ref.~\cite{Altmannshofer:2024hqd}}
\label{Fig:tridents}
\end{figure}

An example diagram of a trident process is given in Fig.~\ref{Fig:tridents} (left); the trident signal considered here is characterized by two long muon tracks in the emulsion coming from the same vertex. The signal muons are required to travel through the FASER$\nu$2 interface tracker, veto station, spectrometer, electromagnetic calorimeter, and an iron block in order to be identified as muons. In contrast, the background events contain at least one charged hadron track. In the case of pions, this can be long, but typically ends to a hadronic interaction either in the tungsten, lead, or iron and can be discarded. However, a charmed hadron may have a short track and decay into a second muon, thus mimicking the trident signature. Fig.~\ref{Fig:tridents} (right) illustrates that near-perfect background rejection, with minimal effect to the signal, is achieved at FASER$\nu$2. This is based on an event analysis relying on reverse tracking between the interface tracker and the emulsion, and restricting the angle between
two outgoing muons to $\theta < 0.1$~rad, parent meson decay length
to $d < 2$~mm, and the number of charged tracks $N_{\rm ch}$ with
momentum $p > 300$~MeV to exactly 2. With these cuts, FASER$\nu$2 has potential for a definitive observation of neutrino tridents with dimuon final states~\cite{Altmannshofer:2024hqd}.

\section{Conclusions}
\label{sec:conclusions}

The first measurements of neutrino interaction cross sections by the FASER Collaboration mark the beginning of the era of using LHC neutrinos for physics. To maximize the physics potential of the LHC and future colliders, it is important to consider the possibilities of an expanded forward neutrino program.
Proposed forward experiments at the LHC have good prospects for e.g. solving the cosmic ray muon excess, probing proton and nuclear PDFs, as well as constraining neutrino NSI and 4-Fermi interactions. Additionally, FASER$\nu$2 has great potential for a conclusive observation of neutrino tridents, probing physics at the EW scale and below. Moreover, synergies can be expected between measurements to be performed at the EIC and at the FPF, and forward neutrino experiments at the FCC could probe proton and nuclear structure at even lower $x$ than is currently possible.

\section*{Acknowledgements}

TM is supported in part by U.S.~National Science Foundation Grants PHY-2111427 and PHY-2210283 and Heising-Simons Foundation Grant 2020-1840.


\medskip

\printbibliography

@article{FASER:2023zcr,
    author = "Abreu, Henso and others",
    collaboration = "FASER",
    title = "{First Direct Observation of Collider Neutrinos with FASER at the LHC}",
    eprint = "2303.14185",
    archivePrefix = "arXiv",
    primaryClass = "hep-ex",
    reportNumber = "CERN-EP-2023-056",
    doi = "10.1103/PhysRevLett.131.031801",
    journal = "Phys. Rev. Lett.",
    volume = "131",
    number = "3",
    pages = "031801",
    year = "2023"
}

@article{SNDLHC:2023pun,
    author = "Albanese, R. and others",
    collaboration = "SND@LHC",
    title = "{Observation of Collider Muon Neutrinos with the SND@LHC Experiment}",
    eprint = "2305.09383",
    archivePrefix = "arXiv",
    primaryClass = "hep-ex",
    reportNumber = "CERN-EP-2023-092",
    doi = "10.1103/PhysRevLett.131.031802",
    journal = "Phys. Rev. Lett.",
    volume = "131",
    number = "3",
    pages = "031802",
    year = "2023"
}

@article{FASER:2024hoe,
    author = "Mammen Abraham, Roshan and others",
    collaboration = "FASER",
    title = "{First Measurement of \ensuremath{\nu}e and \ensuremath{\nu}\ensuremath{\mu} Interaction Cross Sections at the LHC with FASER\textquoteright{}s Emulsion Detector}",
    eprint = "2403.12520",
    archivePrefix = "arXiv",
    primaryClass = "hep-ex",
    reportNumber = "CERN-EP-2024-079",
    doi = "10.1103/PhysRevLett.133.021802",
    journal = "Phys. Rev. Lett.",
    volume = "133",
    number = "2",
    pages = "021802",
    year = "2024"
}

@techreport{Boyd:2882503,
      collaboration = "FASER",
      author        = "Boyd, J and others",
      title         = "{Request to run FASER in Run 4}",
      institution   = "CERN",
      reportNumber  = "CERN-LHCC-2023-009, LHCC-I-039",
      address       = "Geneva",
      year          = "2023",
      url           = "https://cds.cern.ch/record/2882503",
}

@techreport{Abbaneo:2895224,
      collaboration = "SND@LHC",
      author        = "Abbaneo, D and others",
      title         = "{AdvSND, The Advanced Scattering and NeutrinoDetector at
                       High Lumi LHC Letter of Intent}",
      institution   = "CERN",
      reportNumber  = "CERN-LHCC-2024-007, LHCC-I-040",
      address       = "Geneva",
      year          = "2024",
      url           = "https://cds.cern.ch/record/2895224",
      howpublished  = "" 
}

@article{Anchordoqui:2021ghd,
    author = "Anchordoqui, Luis A. and others",
    title = "{The Forward Physics Facility: Sites, experiments, and physics potential}",
    eprint = "2109.10905",
    archivePrefix = "arXiv",
    primaryClass = "hep-ph",
    reportNumber = "BNL-222142-2021-FORE, CERN-PBC-Notes-2021-025, DESY-21-142, DESY-21-142,
  FERMILAB-CONF-21-452-AE-E-ND-PPD-T, KYUSHU-RCAPP-2021-01, LU TP 21-36,
  PITT-PACC-2118, SMU-HEP-21-10, UCI-TR-2021-22, FERMILAB-CONF-21-452-AE-E-ND-PPD-T",
    doi = "10.1016/j.physrep.2022.04.004",
    journal = "Phys. Rept.",
    volume = "968",
    pages = "1--50",
    year = "2022"
}

@article{Adhikary:2024nlv,
    author = "Adhikary, Jyotismita and others",
    title = "{Science and Project Planning for the Forward Physics Facility in Preparation for the 2024-2026 European Particle Physics Strategy Update}",
    eprint = "2411.04175",
    archivePrefix = "arXiv",
    primaryClass = "hep-ex",
    month = "11",
    year = "2024"
}

@article{Feng:2022inv,
    author = "Feng, Jonathan L. and others",
    title = "{The Forward Physics Facility at the High-Luminosity LHC}",
    eprint = "2203.05090",
    archivePrefix = "arXiv",
    primaryClass = "hep-ex",
    reportNumber = "UCI-TR-2022-01, CERN-PBC-Notes-2022-001, INT-PUB-22-006, BONN-TH-2022-04, FERMILAB-PUB-22-094-ND-SCD-T",
    doi = "10.1088/1361-6471/ac865e",
    journal = "J. Phys. G",
    volume = "50",
    number = "3",
    pages = "030501",
    year = "2023"
}

@techreport{Abbaneo:2909524,
      collaboration = "SND@LHC",
      author        = "Abbaneo, D and others",
      title         = "{Addendum to the AdvancedSND LoI}",
      institution   = "CERN",
      reportNumber  = "CERN-LHCC-2024-014, LHCC-I-040-ADD-1",
      address       = "Geneva",
      year          = "2024",
      url           = "https://cds.cern.ch/record/2909524",
}

@article{MammenAbraham:2024gun,
    author = "Mammen Abraham, Roshan and Adhikary, Jyotismita and Feng, Jonathan L. and Fieg, Max and Kling, Felix and Li, Jinmian and Pei, Junle and Rabemananjara, Tanjona R. and Rojo, Juan and Trojanowski, Sebastian",
    title = "{FPF@FCC: Neutrino, QCD, and BSM Physics Opportunities with Far-Forward Experiments at a 100 TeV Proton Collider}",
    eprint = "2409.02163",
    archivePrefix = "arXiv",
    primaryClass = "hep-ph",
    reportNumber = "UCI-TR-2024-13",
    month = "9",
    year = "2024"
}

@article{PhysRevD.108.095020,
    title = {Investigating the fluxes and physics potential of LHC neutrino experiments},
    author = {Kling, Felix and M\"akel\"a, Toni and Trojanowski, Sebastian},
    journal = {Phys. Rev. D},
    volume = {108},
    issue = {9},
    pages = {095020},
    numpages = {18},
    year = {2023},
    month = {Nov},
    publisher = {American Physical Society},
    doi = {10.1103/PhysRevD.108.095020},
    url = {https://link.aps.org/doi/10.1103/PhysRevD.108.095020}
}

@article{Anchordoqui:2022fpn,
    author = "Anchordoqui, Luis A. and Canal, Carlos Garcia and Kling, Felix and Sciutto, Sergio J. and Soriano, Jorge F.",
    title = "{An explanation of the muon puzzle of ultrahigh-energy cosmic rays and the role of the Forward Physics Facility for model improvement}",
    eprint = "2202.03095",
    archivePrefix = "arXiv",
    primaryClass = "hep-ph",
    reportNumber = "DESY-22-021",
    doi = "10.1016/j.jheap.2022.03.004",
    journal = "JHEAp",
    volume = "34",
    pages = "19--32",
    year = "2022"
}

@article{Falkowski:2021bkq,
    author = "Falkowski, Adam and Gonz\'alez-Alonso, Mart\'\i{}n and Kopp, Joachim and Soreq, Yotam and Tabrizi, Zahra",
    title = "{EFT at FASER\ensuremath{\nu}}",
    eprint = "2105.12136",
    archivePrefix = "arXiv",
    primaryClass = "hep-ph",
    doi = "10.1007/JHEP10(2021)086",
    journal = "JHEP",
    volume = "10",
    pages = "086",
    year = "2021"
}

@article{Altmannshofer:2024hqd,
    author = {Altmannshofer, Wolfgang and M\"akel\"a, Toni and Sarkar, Subir and Trojanowski, Sebastian and Xie, Keping and Zhou, Bei},
    title = "{Discovering neutrino tridents at the Large Hadron Collider}",
    eprint = "2406.16803",
    archivePrefix = "arXiv",
    primaryClass = "hep-ph",
    reportNumber = "FERMILAB-PUB-24-0294-T, MSUHEP-24-007",
    doi = "10.1103/PhysRevD.110.072018",
    journal = "Phys. Rev. D",
    volume = "110",
    number = "7",
    pages = "072018",
    year = "2024"
}

@article{PDF4LHCWorkingGroup:2022cjn,
    author = "Ball, Richard D. and others",
    collaboration = "PDF4LHC Working Group",
    title = "{The PDF4LHC21 combination of global PDF fits for the LHC Run III}",
    eprint = "2203.05506",
    archivePrefix = "arXiv",
    primaryClass = "hep-ph",
    reportNumber = "Edinburgh 2021/31, FERMILAB-PUB-22-121-QIS-SCD-T, MSUHEP-22-010,
  Nikhef 2021-033, SMU-HEP-22-01",
    doi = "10.1088/1361-6471/ac7216",
    journal = "J. Phys. G",
    volume = "49",
    number = "8",
    pages = "080501",
    year = "2022"
}

@article{Paukkunen:2014zia,
    author = "Paukkunen, Hannu and Zurita, Pia",
    title = "{PDF reweighting in the Hessian matrix approach}",
    eprint = "1402.6623",
    archivePrefix = "arXiv",
    primaryClass = "hep-ph",
    doi = "10.1007/JHEP12(2014)100",
    journal = "JHEP",
    volume = "12",
    pages = "100",
    year = "2014"
}

@article{Schmidt:2018hvu,
    author = "Schmidt, Carl and Pumplin, Jon and Yuan, C. P. and Yuan, P.",
    title = "{Updating and optimizing error parton distribution function sets in the Hessian approach}",
    eprint = "1806.07950",
    archivePrefix = "arXiv",
    primaryClass = "hep-ph",
    reportNumber = "MSUHEP-18-006",
    doi = "10.1103/PhysRevD.98.094005",
    journal = "Phys. Rev. D",
    volume = "98",
    number = "9",
    pages = "094005",
    year = "2018"
}

@article{AbdulKhalek:2018rok,
    author = "Abdul Khalek, Rabah and Bailey, Shaun and Gao, Jun and Harland-Lang, Lucian and Rojo, Juan",
    title = "{Towards Ultimate Parton Distributions at the High-Luminosity LHC}",
    eprint = "1810.03639",
    archivePrefix = "arXiv",
    primaryClass = "hep-ph",
    reportNumber = "Nikhef/2018-041",
    doi = "10.1140/epjc/s10052-018-6448-y",
    journal = "Eur. Phys. J. C",
    volume = "78",
    number = "11",
    pages = "962",
    year = "2018"
}

@article{HERAFitterdevelopersTeam:2015cre,
    author         = "Camarda, S. and others",
    collaboration  = "HERAFitter developers' team",
    title          = "{QCD analysis of $W$- and $Z$-boson production at 
                       Tevatron}",
    eprint         = "1503.05221",
    archivePrefix  = "arXiv",
    primaryClass   = "hep-ph",
    reportNumber   = "DESY-15-035, FERMILAB-PUB-15-209-PPD, 
                      DESY-REPORT-15-035, DESY Report 15-035",
    doi            = "10.1140/epjc/s10052-015-3655-7",
    journal        = "Eur. Phys. J. C",
    volume         = "75",
    number         = "9",
    pages          = "458",
    year           = "2015"
}

@article{Alekhin:2014irh,
    author = "Alekhin, S. and others",
    title = "{HERAFitter}",
    eprint = "1410.4412",
    archivePrefix = "arXiv",
    primaryClass = "hep-ph",
    reportNumber = "DESY-14-188, DESY-REPORT-14-188, FERMILAB-PUB-14-603-CMS",
    doi = "10.1140/epjc/s10052-015-3480-z",
    journal = "Eur. Phys. J. C",
    volume = "75",
    number = "7",
    pages = "304",
    year = "2015"
}

@article{Bertone:2017tig,
    author         = "Bertone, V. and Botje, M. and Britzger, D. and others",
    title          = "{xFitter} 2.0.0: An Open Source {QCD} Fit Framework",
    booktitle      = "Proceedings, 25th International Workshop on
                      Deep-Inelastic Scattering and Related Topics ({DIS} 2017):
                      Birmingham, {UK}",
    journal        = "PoS",
    volume         = "DIS2017",
    year           = "2018",
    pages          = "203",
    doi            = "10.22323/1.297.0203",
    eprint         = "1709.01151",
    archivePrefix  = "arXiv",
    primaryClass   = "hep-ph",
    SLACcitation   = "%%CITATION = ARXIV:1709.01151;%%"
}

@inproceedings{xFitter:2022zjb,
    author         = "Abdolmaleki, H. and others",
    collaboration  = "xFitter",
    title          = "{xFitter: An Open Source QCD Analysis Framework. A 
                       resource and reference document for the Snowmass 
                       study}",
    eprint         = "2206.12465",
    archivePrefix  = "arXiv",
    primaryClass   = "hep-ph",
    month          = "6",
    year           = "2022"
}

@misc{xFitter:web,
    author         = "xFitter Developers' Team",
    url            = "https://www.xfitter.org/xFitter/",
    howpublished   = ""
}

@article{Eskola:2021nhw,
    author = "Eskola, Kari J. and Paakkinen, Petja and Paukkunen, Hannu and Salgado, Carlos A.",
    title = "{EPPS21: a global QCD analysis of nuclear PDFs}",
    eprint = "2112.12462",
    archivePrefix = "arXiv",
    primaryClass = "hep-ph",
    doi = "10.1140/epjc/s10052-022-10359-0",
    journal = "Eur. Phys. J. C",
    volume = "82",
    number = "5",
    pages = "413",
    year = "2022"
}

@article{Cruz-Martinez:2023sdv,
    author = {Cruz-Martinez, Juan M. and Fieg, Max and Giani, Tommaso and Krack, Peter and M\"akel\"a, Toni and Rabemananjara, Tanjona R. and Rojo, Juan},
    title = "{The LHC as a Neutrino-Ion Collider}",
    eprint = "2309.09581",
    archivePrefix = "arXiv",
    primaryClass = "hep-ph",
    reportNumber = "Nikhef-2023-009, CERN-TH-2023-165",
    doi = "10.1140/epjc/s10052-024-12665-1",
    journal = "Eur. Phys. J. C",
    volume = "84",
    number = "4",
    pages = "369",
    year = "2024"
}

@article{PierreAuger:2014ucz,
    author = "Aab, Alexander and others",
    collaboration = "Pierre Auger",
    title = "{Muons in Air Showers at the Pierre Auger Observatory: Mean Number in Highly Inclined Events}",
    eprint = "1408.1421",
    archivePrefix = "arXiv",
    primaryClass = "astro-ph.HE",
    reportNumber = "FERMILAB-PUB-14-290-AD-AE-E-TD",
    doi = "10.1103/PhysRevD.91.032003",
    journal = "Phys. Rev. D",
    volume = "91",
    number = "3",
    pages = "032003",
    year = "2015",
    note = "[Erratum: Phys.Rev.D 91, 059901 (2015)]"
}

@article{PierreAuger:2016nfk,
    author = "Aab, Alexander and others",
    collaboration = "Pierre Auger",
    title = "{Testing Hadronic Interactions at Ultrahigh Energies with Air Showers Measured by the Pierre Auger Observatory}",
    eprint = "1610.08509",
    archivePrefix = "arXiv",
    primaryClass = "hep-ex",
    reportNumber = "FERMILAB-PUB-16-504-AD-AE-CD-TD",
    doi = "10.1103/PhysRevLett.117.192001",
    journal = "Phys. Rev. Lett.",
    volume = "117",
    number = "19",
    pages = "192001",
    year = "2016"
}

@article{PierreAuger:2021qsd,
    author = "Aab, Alexander and others",
    collaboration = "Pierre Auger",
    title = "{Measurement of the Fluctuations in the Number of Muons in Extensive Air Showers with the Pierre Auger Observatory}",
    eprint = "2102.07797",
    archivePrefix = "arXiv",
    primaryClass = "hep-ex",
    reportNumber = "FERMILAB-PUB-21-202-AD-AE-SCD-TD, FERMILAB-PUB-21-202-AD-AE-SCD-TD",
    doi = "10.1103/PhysRevLett.126.152002",
    journal = "Phys. Rev. Lett.",
    volume = "126",
    number = "15",
    pages = "152002",
    year = "2021"
}

@article{CHARM-II:1990dvf,
    author = "Geiregat, D. and others",
    collaboration = "CHARM-II",
    title = "{First observation of neutrino trident production}",
    reportNumber = "CERN-EP-90-75",
    doi = "10.1016/0370-2693(90)90146-W",
    journal = "Phys. Lett. B",
    volume = "245",
    pages = "271--275",
    year = "1990"
}

@article{CCFR:1991lpl,
    author = "Mishra, S. R. and others",
    collaboration = "CCFR",
    title = "{Neutrino Tridents and W Z Interference}",
    reportNumber = "NEVIS-1437, FERMILAB-PUB-91-390",
    doi = "10.1103/PhysRevLett.66.3117",
    journal = "Phys. Rev. Lett.",
    volume = "66",
    pages = "3117--3120",
    year = "1991"
}

@article{NuTeV:1999wlw,
    author = "Adams, T. and others",
    collaboration = "NuTeV",
    title = "{Evidence for diffractive charm production in muon-neutrino Fe and anti-muon-neutrino Fe scattering at the Tevatron}",
    eprint = "hep-ex/9909041",
    archivePrefix = "arXiv",
    reportNumber = "FERMILAB-PUB-99-269-E",
    doi = "10.1103/PhysRevD.61.092001",
    journal = "Phys. Rev. D",
    volume = "61",
    pages = "092001",
    year = "2000"
}

\end{document}